\documentstyle[preprint,aps]{revtex}

\begin{document}
\draft
\newcommand{\ch}{\mbox{$\chi^{(2)\;}$}}
\newcommand{\cw}{\mbox{$\chi^{(2)}(\omega)\;$}}
\newcommand{\cwt}{\mbox{$\chi^{(2)}(\omega ,T)\;$}}
\newcommand{\cwtel}{\mbox{$\chi^{(2)}(\omega ,T_{el})\;$}}
\newcommand{\cz}{\mbox{$\chi^{(2)}_{zzz}(\omega ,T)\;$}}
\newcommand{\iwt}{\mbox{$I^{(2)}(\omega ,T)\;$}}
\newcommand{\iwtel}{\mbox{$I^{(2)}(\omega ,T_{el})\;$}}
\newcommand{\epsw}{\mbox{$\epsilon(\omega ) \;$}}
\newcommand{\ewt}{\mbox{$\epsilon(\omega ,T) \;$}}

%
\title{Theory for the Dependence of Optical Second Harmonic Generation
  Intensity on Non-equilibrium Electron Temperatures at Metal Surfaces }

\author{T. A. Luce, W. H\"ubner, and K. H. Bennemann}
\address{Institute for Theoretical Physics, Freie Universit\"at 
Berlin, Arnimallee 14, D-14195 Berlin, Germany}
\date{\today}
\maketitle
\begin{abstract}
We present a theory for the electron-temperature
dependence $T_{el}$ of optical second harmonic generation (SHG). Such an
analysis is required to 
study the dynamics of metallic systems with many hot electrons not at
equilibrium with the lattice.
Using a tight-binding theory for the nonlinear susceptibility \cwtel
and the Fresnel coefficients we present results for the SHG intensity
\iwtel for a Cu monolayer. In the case of linear optical response we
find that the intensity will decrease monotonously
for increasing $T_{el}$. In
agreement with experiment we find a frequency range where \iwtel
may be enhanced or reduced depending on electron temperature.
Note, \cwtel rather than the Fresnel coefficients determines essentially the
temperature dependence. Our theory yields also that SHG probes effects
due to hot electrons more sensitively than linear optics. We also
discuss the $T_{el}$-dependence of SHG for Au and Ag.

\end{abstract}

\pacs{73.20.At, 78.47.+p, 78.66.-w}

\newpage

\section{Introduction}
Due to its surface sensitivity, the nonlinear optical response has
become a powerful probe for investigating the electronic
structure of surfaces, interfaces, thin films, and multilayers.
Recently, a combination of linear and nonlinear experiments
\cite{hohlf95} has been performed 
exploiting the time dependence of SHG. Thus, effects of hot electrons
not at equilibrium with the lattice and their changes in time can be analyzed.
This opens a new route to investigate the dynamics of the system
during relaxation to the equilibrium state. Note, different
electronic temperatures of hot electrons not at equilibrium with the
lattice but among themselves result from varying the light
irradiation. If intense and short laser pulses in the range of 100 fs to  1 ps 
are used for the SHG experiment, only the electrons will quickly 
thermalize (even in the approximation of Fermi liquid theory), since 
the slow electron-phonon coupling does not come into
play \cite{stampfli90,stampfli92}. Only later, at
times of the order of several picoseconds the electrons will heat up
the lattice. Thus, at short times the temperature dependence of SHG is
essentially due to the varying non-equilibrium electronic temperature,
 which may be considerably different from the equilibrium
temperature at later times, when electrons and lattice are at equilibrium.

In this paper, we present a theory for the time-resolved optical SHG
response, in particular its dependence on non-equilibrium electronic 
temperature. We present results for the electron temperature
dependence of the SHG yield at noble metal surfaces using an
electronic theory for the nonlinear response. The metallic surface is
modeled by a freestanding Cu monolayer. However, 
the model allows already to identify characteristic features of this
time-dependent nonlinear optical response to be expected also for the
surface of bulk Cu and for other metals like Ag and Au and to explain recent
experimental results for Cu and Au \cite{hohlf95b}. Our analysis allows
to identify the essential origin of the electron-temperature
dependence of the nonlinear optical response. Furthermore our studies
explain the different behavior of linear and nonlinear optics on
non-equilibrium electronic temperatures. In general, our
theory is of interest for the dynamics of nonequilibrium electronic
systems. 

In section II, we describe details of our theory. In section
 III, we present results of our calculation of the nonlinear
response for several values of the electron temperature and compare with
experiments. The conclusions in section IV point out possible
general features of our model calculation and explain why the
important temperature dependence results from \cwt and not from
the Fresnel coefficients. In the appendix some details of our
calculation are presented.

\section{Theory}
We calculate the SHG intensity \iwt using an electronic theory. 
Then the SHG yield for $p$-polarization within the electric dipole
approximation is~\cite{hub_bohm94}

\begin{eqnarray}
I(p-SH)\;&&=\;\mid (2i\frac{\omega}{c})\mid^{2}\mid
E_{0}(\omega)\mid^{4}\mid A_{p}((F_{c}\chi^{(2)}_{xzx}
2f_{c}f_{s}+N^{2}F_{s}(\chi^{(2)}_{zxx}f_{c}^{2}+\chi^{(2)}_{zzz}f_{s}^{2}))
t_{p}^{2}\cos^{2}\varphi\;\nonumber \\
&&+\;
N^{2}F_{s}\chi^{(2)}_{zxx}t_{s}^{2}\sin^{2}\varphi)\mid^{2}\;.
\end{eqnarray}
Here, $\varphi $ denotes the angle of polarization of the incident light,
$f_{c,s}$ the Fresnel coefficients and $t_{s,p}$ 
the linear transmission
coefficients, $ n \;=\; \sqrt{\epsilon(\omega)}\;$ and 
 $\; N \;=\; \sqrt{\epsilon(2\omega)} \; $ 
the complex indices of refraction at frequencies $\omega $
and 2$\omega $. Furthermore, $f_{s}\;=\;\sin\theta/n$ and
 $f_{c}\;=\;\sqrt{1-f_{s}^{2}}$
for the fundamental frequency $ \omega $,  
$F_{s}\;=\;\sin\Theta/N$ and $F_{c}\;=\;\sqrt{1-F_{s}^{2}}$ for
the doubled frequency, where $\theta $ and $\Theta $ denote the angle of 
incidence and the angle of reflection of the SHG light, respectively.
The linear transmission coefficients are given 
by~\cite{klausdiss,sipe}
\begin{equation}
t_{p}\;=\;\frac{2\cos\theta}{n\cos\theta+f_{c}}\;,
\;\;t_{s}\;=\;\frac{2\cos\theta}{\cos\theta + nf_{c}}
\;,\;\;T_{p}\;=\;\frac{2\cos\Theta}{N\cos\Theta+F_{c}}\;.
\end{equation}
The amplitude $A_{p}$ in Eq. (1) is
\[
A_{p}=\frac{2\pi T_{p}}{\cos\Theta}\;.
\]
Note, the nonlinear susceptibility tensor
$\chi^{(2)}$ is material specific and so are the linear dielectric 
function $\epsilon (\omega)$
and the indices of refraction $n$ and $N$ via the electronic 
bandstructure.

The contributions to \iwt due to the Fresnel factors $F_c$, $F_s$,
$f_c$, and $f_s$ and the transmission coefficients $T_p$, $t_p$, and
$t_s$ and \cwt are all
temperature dependent. In both cases the temperature dependence arises
from the Fermi functions $f(E,T)$ which, due to the many hot electrons
resulting from the light irradiation, have to be taken at
considerably elevated electron temperatures. Note, we limit ourselves to
the time regime, where the electrons have already 
thermalized, but 
electron-phonon coupling has not become really effective.
By taking both temperature dependences into account it becomes
possible to decide theoretically whether \cwt or the Fresnel and
transmission coefficients cause the essential
temperature dependence of the SHG intensities.
Since it is known from theory and also experiment that the $\chi^{(2)}_{zzz}$ tensor element dominates over $\chi^{(2)}_{xzx}$ and
$\chi^{(2)}_{zxx}$~\cite{hub_bohm94},
we restrict our calculation to this single element of the nonlinear
susceptibility for the SHG yield. Then approximately
\begin{equation}
I(p-SH)\;=\;\mid (2i\frac{\omega}{c})\mid^{2}\mid
E_{0}(\omega)\mid^{4}\mid A_{p}N^{2}F_{s}\chi^{(2)}_{zzz}f_{s}^{2}
t_{p}^{2}\cos^{2}\varphi\mid^{2} \; .
\end{equation}

For the calculation of \ewt and 
\cz, we use eigenvalues $E_{{\bf k},l,\sigma }$ from a 
simple quadratic, freestanding Cu monolayer.
The nearest-neighbor-distance equals $3.61/\sqrt{2}$ \AA ,
according to the nearest-neighbor-distance of Cu bulk.
The bandstructure
involves five $d$-bands and four plane waves and has been 
obtained \cite{ehren-hodges} within the
combined interpolation scheme (CIS).
The $d$-bands are parameterized in terms of 
Fletcher-Wohlfahrt parameters~\cite{fletcher-wohl,fletcher}, 
their values are given in 
Table 1. The parameters  
of the $d$-bands were evaluated by a fit to an
{\em ab initio} LAPW band-structure calculation by Krakauer,
Posternak, and Freeman \cite{freeman79}. Furthermore, to get also the
correct onset of 
interband transitions we shifted correspondingly the $d$ states
\cite{weinb94}. The energy eigenvalues were calculated for 1861 k-points in the 
irreducible part of the Brillouin zone (1/8 of the whole Brillouin
zone). The resulting band structure shown in Fig. 5 should be
thus representative for the surface layer of bulk Cu.

The calculation of the dielectric function \ewt is performed 
by including both the intra- and interband electronic
transitions. Neglecting the $q$-dependence of $\epsilon$, we take
\cite{haug84,pusto93} 

\begin{equation}
    \epsilon{(\omega,T)} = 
    1 - \frac{\omega_{pl}^2}{\omega(\omega + i/{\tau_{pl}})} -
\frac{4\pi e^{2} }{\Omega}\;
         \sum_{{\bf k},l,l^{\prime },{\sigma}} 
    {{\it M}^2  
        \times \frac{f(E_{{\bf k}l^{\prime }\sigma },T)-
f(E_{{\bf k}l\sigma },T)} 
        {E_{{\bf k}l^{\prime }\sigma }-E_{{\bf k}l\sigma }
        -\hbar \omega +i\hbar \alpha_{1}} }  \; ,
\label{eps}
\end{equation}
where ${\it M}$ are the dipole matrix elements, 
$f(E,T)$ are the Fermi functions, $ \omega $ is the fundamental frequency, 
$ \Omega$ is the unit area, $ \alpha_1= 0.1 \; $ eV is the Lorentzian broadening,
$ \tau_{pl}$ the 
Drude relaxation time and 
$ \omega_{pl}\; = \; {4 \pi n_c e^2}/{{m^{\ast}}_c}$ 
is the plasma frequency with $n_c$ the electron density 
and $m^{*}_c$ 
the effective electron mass. Both $  \omega_{pl}$ and 
$\tau_{pl}$ are fitted to a dielectric 
function using literature values~\cite{johnchrist72} 
at low fundamental energies, 
where no interband transitions occur. We use 
  $ \hbar \omega_{pl} $ = 9 eV and  
$\tau_{pl}/\hbar$ = 8.5 ${\rm eV}^{-1}$.

To simplify our calculation, we assume constant matrix elements $M$, 
which fit to the dielectric function
$\epsilon(\omega, T = 0 K)$ \cite{pusto93}. This approximation 
is reasonable because 
the {\bf k}-dependence of the matrix elements is expected to
become less important in two dimensions due to the shrinking of the
$d$-band width for the reduced coordination number and also due to the
occurrence of additional allowed optical transitions.
Additionally, as will be shown later, the main contribution of the SHG 
intensity has its origin in $d \rightarrow d \rightarrow s $
transitions, so the matrix elements give a simple prefactor to the
sum and cancel, when only intensity differences are considered.
Once \ewt is calculated, it is straightforward to
get the Fresnel factors and transmission factors from Eq. (2).
\\
For evaluating the second-order susceptibility, we employ the 
microscopic theory developed in Ref. \cite{hubner90}. Thus, the tensor
element $\chi^{(2)}_{zzz}$ is given by

 \begin{eqnarray}
   \chi^{(2)}_{zzz}(\omega,T)&=&
    \frac{e^{3}}{\Omega}\sum_{{\bf k},l,l^{\prime },l^{\prime \prime },\sigma}
     ({\it M}_{z})^3   \Bigg\{ \frac 
    {\frac{f(E_{{\bf k},l^{\prime \prime }\sigma},T)-
           f(E_{{\bf k},l^{\prime }\sigma },T)} 
         {E_{{\bf k},l^{\prime \prime }\sigma }-
          E_{{\bf k},l^{\prime }\sigma }
          -\hbar \omega + i\hbar \alpha_{1}} 
    -\frac{f(E_{{\bf k},l^{\prime }\sigma },T)-
           f(E_{{\bf k}l\sigma },T)}
         {E_{{\bf k},l^{\prime }\sigma }-
          E_{{\bf k}l\sigma }-\hbar \omega +i\hbar \alpha_{1}} }
    {E_{{\bf k},l^{\prime \prime }\sigma }-
     E_{{\bf k}l\sigma }-2\hbar \omega +i2\hbar \alpha_{1}} 
      \Bigg\}  \; .
\end{eqnarray}
Note, for experiments not detecting absolute intensities (as is usually
the case), the
prefactor in this formula is of no further importance for comparison
with experimental data.
As in the calculation of \epsw, we neglect the {\bf{k}}-dependence 
of the matrix elements, 
taking the values for ${\it M}_{z}$ from the $\epsilon(\omega)$-fit. 
We only take into account the dominant $\chi^{(2)}_{zzz}$-tensor
element and neglect the contributions of the other tensor elements, as
 pointed out earlier. This completes then the theory for the
 determination of the SHG intensity \iwt with the two essential inputs
 \ewt and \cwt.

\section{Results}

The most important results of our calculations are presented in
Figs. 1-4. For these results we used an
electronic structure which is shown in Fig. 5 and which was obtained
using inputs discussed in section II.
In Fig. 1 we present results for the change 
\[ \Delta I^{(2)}(\omega) = \frac{I^{(2)}(\omega,T_{el})-
  I^{(2)}(\omega,300 K)}{I^{(2)}(\omega,300 K)}  \] of
the SHG yield as
a function of frequency for different non-equilibrium electronic
temperatures $T_{el}$. In particular, in the inset of Fig. 1 we show for the
frequency range from $0.4$ to $1.5$ eV, that $I^{(2)}(\omega )$ may be
reduced or enhanced due to increasing non-equilibrium temperature for
the electrons. 
In order to demonstrate what causes essentially the dependence on the
electronic temperature, namely \cwt or the Fresnel
coefficients, we present in Fig. 2 results for 
\[ \Delta \chi^{(2)}(\omega  ,T) =\frac{\chi^{(2)}(\omega  ,T_{el}) -
\chi^{(2)}(\omega  ,300 K)}{ \chi^{(2)}(\omega  ,300 K)} \;, \] and in
Fig. 3 results for the difference \[ \Delta \frac{I^{(2)}}{|\chi^{(2)}|^2}
= \frac{I^{(2)}(\omega ,T_{el})/{|\chi^{(2)}(\omega ,T_{el})|^2} -
I^{(2)}(\omega ,300 K)/ {|\chi^{(2)}(\omega ,300 K)|^2}}
{ I^{(2)}(\omega ,300 K)/{|\chi^{(2)}(\omega ,300 K)|^2}}  \;. \]
Note, in Fig. 3 we present
results for $I^{(2)}(\omega ,T)$ using a frequency and temperature independent 
susceptibility $\chi^{(2)}$. The comparison of the results in Figs. 2 
and 3 show clearly that \iwt results essentially from
\cwt. Furthermore, from comparing the results in Figs. 1 and 3
together with those in 2 and 3, we note that the crossover behavior
with respect to the 
temperature dependence is only present in the case of the nonlinear
response \iwt, but not in the Fresnel and transmission coefficients.

In Fig. 4 we present results for $I^{(2)}(\omega ,T=2000 K)$ as a function of frequency. The
different curves (b), (c), (d), and (e) refer to cases where we
artificially excluded certain optical transitions from contributing to
\cwt . Thus, we are able to demonstrate which transitions are of
particular significance for the temperature dependence of \iwt . While
curve (a) still includes all transitions, in the case of curve (b) we
excluded contributions to \cwt , where the initial electronic state is
not in the $d$-band. In the case of curve (c) we excluded contributions
to \cwt , where the initial state is not in the $d$-band or the
intermediate state is in the $d$-band. Curve (d) results when the
initial and the intermediate states are in
the $d$-band, and in curve (e) we excluded transitions, where the initial
state is in the $d$-band or the intermediate state is not in the
$d$-band.
Similar results are obtained for 300 K and 6000 K.

For a detailed analysis of the results presented in Figs. 1 - 4 we
like to make the following remarks.
Investigating the relative changes of SHG intensity with temperature 
and its contributions due to the Fresnel factors and the nonlinear
susceptibility \cwt , we observe in Figs. 1, 2, and 3, 
that the shape of the 
dependence on the fundamental frequency is given by the 
second-order susceptibility \cwt for fundamental energies below 3 eV.
In the range between 1.2 eV and 3 eV, the SHG intensity decreases
with increasing temperature.  Most important is the result that indeed
an energy window at a fundamental energy of 1.1 eV exists, where a small
temperature increase (300 K to 2000 K) results in a SHG intensity
increase, but 
a stronger temperature increase (300 K to 6000 K) results in a 
rapid SHG intensity decrease. From Figs. 2 and 3 we conclude that \cwt
is responsible for this. This clarifies then the physical origin of
the temperature dependence of \iwt observed on Cu polycrystalline surfaces~\cite{hohlf95b}, 
where in  2 eV photon energy pump--probe SHG experiments 
on Cu surfaces for weak 
pump pulses an increase of the SHG intensity has been observed, but
a decrease when the pump pulse becomes stronger and thus the electron 
temperature higher. 

Note, it is remarkable that our model
bandstructure yields already such a fair agreement with the
experimental results. This is so, since the temperature dependence of
\cwt and \iwt is strongly governed by the correct position of the
upper $d$ band edge and also because the important features of the
bandstructure are already simulated correctly by our model. Furthermore, SHG
probes the surface layer only, so our bandstructure is a fair
approximation for the experimental situation.
In view of this, we may also use with proper changes the model in
Fig. 5 to describe the SHG response of Au and Ag. For this we change
the onset of $d$-band transitions $\Delta$ to $\Delta = 3.95$ eV in
the case of Ag and to $\Delta = 2.3$ eV in the case of Au, since this
is expected to cause the most important differences, s. Fig. 5. The
other energy levels shift accordingly. 

Since the ``mismatch'' between $(E_d$ - $E_F)$ and the photon energy
$\hbar \omega$ in Au is 
by 230 meV larger than in Cu (Cu bulk: $E_d$ - $E_F = 2.15$ eV), 
one may simulate in our calculation the Au surface by
using approximately the SHG spectrum of the Cu monolayer, but
considering a photon 
energy which is 230 meV below the respective energy value for the Cu
monolayer. While for Cu we found the crossover behavior 
at 1.1 eV, we have to take for Au a fundamental photon energy of 0.87 eV.
Then our results presented in Fig. 1 yield an increase of SHG for all
temperatures. The increase for 6000 K is higher than that for 2000
K. These results seem in reasonable agreement with experiments for a
Au polycrystalline surface, where a monotonic increase of the SHG
yield for both temperatures has been observed \cite{hohlf95b}.
Thus, our electronic theory is able to explain the ultrafast
electronic relaxation process on both Cu and Au surfaces. From our
direct calculation of the dielectric functions (see Fig. 6) we find
that the temperature dependence of $I^{(2)}$ is mainly caused by \cwt.
At a frequency $\hbar \omega = 0.9 $ eV we find that the  
\cwt -dependence on the electron temperature
begins to saturate for higher temperature, as is observed in the experiment 
\cite{hohlf95b,hohlf95c}.  
Note, the SHG response of Ag may be modeled similarly as has been
described for Au. Further remarks on this are given later. 

Regarding the origin of the temperature dependence of the SHG
contribution (on transitions), it is necessary to investigate the 
contribution of the $d$-band electrons to
\cz and its influence on the SHG yield in some
detail. Results of this analysis are shown in Fig. 4. To interpret the
results, one has to be careful not to neglect the interferences
between the various terms in Eq. (5). 
Therefore, calculating \cz and the SHG intensity
for just a few transitions, neglecting all others, is of little value,
since then the interferences between the various complex quantities
are almost completely neglected.
Thus, we calculated the SHG intensity with a \cz, Authors: T. A. Luce, W. H"ubner, and K. H. Bennemann

where just some transitions with a specific combination of the three
states necessary for a nonlinear transition were neglected.
States were identified as $d$-states if their energy eigenvalue was 
between -5.25 and -1.45 eV below $E_F$. Note, here we neglected the
symmetry of the states.
Apparently, we find that the initial state is almost always a $d$
state, since both intensities nearly coincide. 
The SHG intensity \iwt from transitions $d \rightarrow (s,p) \rightarrow
(d,s,p)$, where the initial state is in the $d$-band and 
the intermediate state is not (curve (c)), nearly
vanishes. The ratio of
the maximum values of the SHG intensity and the total SHG
intensity $I^{(2)}(d \rightarrow (s,p) \rightarrow
(d,s,p))/I^{(2)}(total)$ is equal to $1/90$. Of course, both
intensities are taken at the same frequency. This 
shows that most transitions contributing to the SHG intensity have the
initial and intermediate state in the $d$-band. This is confirmed by the
respective dotted curve (d). Due to interference of the various
interband
transitions
in \cz ,
this curve and the total SHG intensity yield differ at the maximum 
by a factor of 3. Comparing this ratio $I^{(2)}(d \rightarrow d \rightarrow
(d,s,p))/I^{(2)}(total)\;=\;3$ with the ratio 
$I^{(2)}(d \rightarrow (s,p) \rightarrow (d,s,p))/I^{(2)}(total)$, it is 
reasonable to assume that only the transitions with the initial and
the intermediate state in the $d$-band contribute significantly to the
SHG yield.  

From curve (e) in Fig. 4 it can be seen that considerable SHG intensity is created by 
$(s,p) \rightarrow d \rightarrow (s,p,d)$ transitions. 
This is readily understood if one notes that all terms in 
the sum of $\chi^{(2)}_{zzz}(\omega ,T)$,  Eq. (5), consist of differences

 \begin{eqnarray}
 \frac {1}{E_3-E_1 -2\hbar \omega + 2i\hbar \alpha_{1}} 
    \Bigg\{\frac{f(E_3, T)- f(E_2, T)}{E_3-E_2 -\hbar \omega + i\hbar \alpha_{1}} 
    -\frac{f(E_2, T)- f(E_1, T)}{E_2-E_1 -\hbar \omega + i\hbar
      \alpha_{1}}\Bigg\}\; .
\end{eqnarray}
Here, $E_1$, $E_2$ and  $E_3$ denote the initial, intermediate, and
final state in the considered transition, respectively. 
This difference will give a contribution even if one of the terms in
curly brackets is
zero, so the intensity $I^{(2)}(s,p \rightarrow d \rightarrow s,p,d)$
mainly results from the first term.

Compared with these transitions, all other combinations
of states contribute only a small amount to the SHG. Even at
low energies ($\hbar \omega \sim$ 1 eV), the probability that all
three states of a nonlinear transition are $d$-states 
is rather small. Since the main contribution to the SHG intensity
results from transitions with the two lower states in the $d$-band,
nearly all contributing terms have the corresponding dipole matrix 
element product  
$\langle d | z | d \rangle \langle d | z | (s,p) \rangle 
\langle (s,p) | z | d \rangle $. This can be handled as a prefactor. 
In view of this result, our simplification of 
using one value for all dipole matrix elements is justified.

In order to avoid confusion, one should note that \nolinebreak our
\nolinebreak \mbox{calculation} of
\iwt demonstrates that the mismatch energy $ \delta = $ $ E_F - E_d -
\hbar\omega $ is crucial for the $T$-dependence of the SHG yield and
gives for a fixed photon energy $\hbar\omega $ the following results:   
First, if $\delta < \delta_{Cu}$, where $\delta_{Cu}$ refers to the
value where the reflectivity of Cu saturates, the temperature
dependence of the Fresnel factors and transmission coefficients is
sufficient to describe \iwt. In the range $\delta_{Cu} < \delta <
\delta_{Au}$, \iwt is essentially due to \cwt, and for $\delta > \delta_{Au}$, \iwt results from \cwt for both Cu and Au. This is
consistent with experimental observations regarding the dependence of
the reflectivity on $\delta $.

In summary, our results for \iwt suggest that the temperature
dependence is essentially due to \cwt. In particular, \cwt is
responsible for the nonmonotonous temperature dependence \iwt . 

\section{Conclusions}

We calculated the dependence of the SHG yield on electron
temperatures for electrons not at equilibrium with the lattice, caused by light
irradiation for a Cu surface and for a Ag and a Au surface.
We find that for energies below 3.2 eV the temperature effects result
mainly from \cwt and not from the Fresnel coefficients. In particular,
our calculation yield that in the frequency range between 0.8 
and 1.1 eV, the SHG intensity even may increase
with increasing electron temperature. This is the case
for Cu at low light intensities and for Au for all light intensities
that cause no damage. Note,
for Au we use the same bandstructure and smaller light frequencies in order
to study the same electronic transitions as for Cu. Clearly, \iwt
depends on the position of the $d$-band with respect to the Fermi-energy. 
The interesting effect that mainly \cwt causes the temperature
dependence comes about due to extra two photon absorption processes
making use
of the high $d$-band density of states and occurring only at elevated
electronic temperatures. Note, however, that only due to the interference
of these transitions with others a considerable temperature dependence
comes about. The situation is illustrated in Fig. 7. Note, without the
effects due to the indicated transitions, the nonmonotonic temperature
dependence would be absent. The figure illustrates also the difference
between the temperature dependence in nonlinear and linear response,
where for the latter only $d$-states as initial
states are possible.

As already remarked the essential features of our results should be valid
independent of 
our model calculations. This is supported by the fact that we already
find such a fair agreement with the experiments by Hohlfeld, Conrad
and Matthias \cite{hohlf95b}. 
Our theory shows clearly how \iwt depends on the electronic structure
and thus what can be
expected for other noble metals and transition metals. For example,
for Ag ($E_F - E_d = 3.98 $ eV for bulk Ag) we would expect, using the
previously discussed argument, a similar behavior for the linear and
nonlinear response in the frequency range from 0.8 to 1.1 eV in view of
Fig. 7, while in the frequency range 4 eV a decrease of the SHG yield 
due to the very small negative mismatch is expected. For Ag and for this
mismatch the SHG intensity \iwt should always decrease as a function
of $T$.
Also for Fe, in addition to the changes due to the
electronic structure (position of the $d$-band) we expect interesting
non-equilibrium temperature effects due to the additional dependence of the
magnetization {\bf M} on $T_{el}$, since $I^{(2)}$ depends on $T_{el}$ and ${\bf M}(T_{el})$.

Regarding the time dependence of our results, for larger times the
electron-phonon coupling will
become more effective and cause the decrease of the electronic
temperature via an energy transfer from the hot electrons to the
lattice. The resulting equilibrium temperature will be only a few
hundred degrees higher than before the light irradiation. Then, \iwt will be
similar to $I^{(2)}(\omega, 300 K)$, which is given in the appendix. Such
temperature dependences should be compared with those obtained for
systems due to usual equilibrium thermodynamical effects.
It is straightforward to extend our calculation to thicker films
consisting of several atomic layers. Such calculations are in progress
\cite{hue_luce96}.

As a resum\'{e}, our studies show that SHG can be used to study 
the dynamics of excited hot electrons in crystals. 
Note, in this paper we considered the time window
where the electrons are far from equilibrium and have their own
temperature different from the lattice temperature. Obviously, using
SHG for studying the time evolution of the electronic system far from
equilibrium offers new perspectives, in particular for studying magnetism.

\nonumber
\[{\mbox \rm {\bf ACKNOWLEDGMENT}}\]

We gratefully acknowledge many discussions with
Prof. E. Matthias and J. Hohlfeld stimulating this work.

\nonumber
\[{\mbox \rm {\bf APPENDIX}}\]

In this appendix we present some details of our calculations.
In Fig. 5 we show results for our band-structure calculation,
evaluated with the parameters fitted to the {\em ab initio}
bandstructure. Our CIS calculation is in good agreement 
with the {\em ab initio} calculations. The
resulting density of states (DOS) in the inset of Fig. 6 gives a $d$-band edge
approximately 1.45 eV below the Fermi-energy $E_f$. 
The total $d$-band width is 3.05 eV.
$d$-band transitions with lowest energy are
possible at energies around 2.1 eV (near the $\bar{X}$-point).
The constant values of the matrix elements are $M_{z} = 5.4 *
10^{-10}$ m. The numerical value was obtained by fitting the maximum values of 
the dielectric
function to literature values~\cite{johnchrist72}. From atomic orbitals, one
would expect values about
$0.01*10^{-10}$ m for the $\langle p | z | d \rangle$ 
matrix elements~\cite{hubner90}. This shows that the atomic 
orbitals give the correct symmetry of the wave functions, but the 
absolute values differ considerably due to the 
periodicity of the monolayer wave functions which 
delocalizes the $d$-electrons.

The real and imaginary parts of the dielectric function for the 
Cu monolayer as a function of the incident photon energy are 
plotted in Fig. 6 for three different electronic temperatures (300
K, 2000 K, and 6000 K). 
These temperatures are chosen as realistic estimates of the
electron temperature caused by experimental pump--probe pulses.
 
The influence of the plasma frequency $\omega_{pl}$
at low frequencies leads to a strong drop (increase) of the 
real (imaginary) part of $\epsilon (\omega) $ upon lowering the
frequency $\omega$. A variation of the
Drude relaxation time $\tau_{pl}$ causes only minor changes \cite{footnote1}.
If $\omega_{pl}$ is reduced, the minimum of Im \{$\epsilon$\} increases
in value and moves to lower energies.
The increase of Im \{$\epsilon$\}
at 2 eV is caused by transitions from the $d$-band
to unoccupied states at $E_f$ at this energy. Due to the 
Lorentzian broadening introduced by the 
calculation, the increase of Im \{$\epsilon$\} is not as steep as expected
from optical measurements. Rising the electron temperature
results in a flattening of the curves for Re \{$\epsilon$\} and 
Im \{$\epsilon$\}.
The minimum of Im \{$\epsilon$\}
becomes less pronounced for higher temperatures. Since there are
more vacant
states below $E_f$ at higher temperatures, interband transitions
occur at lower energies, thus weakening the rise of Im \{$\epsilon$\} at 2
eV.
It is to be expected that the main temperature effects are observable
at photon energies $\omega$ with either $\hbar\omega \approx E_f - E_d $ or 
$2\hbar\omega \approx E_f - E_d $, depending on the influence of the Fresnel factors or
the nonlinear susceptibility on the temperature dependence. Then the
transitions are strongly influenced by the effects which a temperature
increase induces on the occupation of states at $E_f$.

In order to trace back the origin of this effect we separately
calculate the contribution of the Fresnel factors and nonlinear 
contributions to the SHG yield. To
obtain the contributions of the Fresnel factors we calculate the SHG intensity
$I(2\omega)$ from Eq. (3) with $\chi^{(2)}_{zzz}$ set to unity, thus 
including the interaction of the incident light and the 
frequency-doubled light with the monolayer. For
the nonlinear contributions we compute \cz .
Inserting the dielectric function from Eq. (4) and  \cz
from Eq. (5) in Eq. (3), in Fig. 8 we show the 
calculated SHG intensity
\iwt for the three different electronic temperatures 
(300 K, 2000 K, 6000 K). 
The SHG intensity rises strongly at photon energies around 2 eV and peaks
at approximately 2.8 eV. Then it drops without exhibiting further peaks. 
For increasing temperature, the intensity maximum decreases and 
moves to slightly higher photon energies.
From Fig. 9 we
find a similar shape of the \cz-plots, with maxima
shifted to photon energies around 1.9 eV. From Eq. (5) it becomes
clear that even at low energies in the range of 1 eV contributions to
\cz are to be expected due to $2\hbar\omega$-resonances
of the form 
\[ \frac{1}{E_{{\bf k}+2{\bf q},l^{\prime \prime }\sigma }-
     E_{{\bf k}l\sigma }-2\hbar \omega +i2\hbar \alpha_{1}} \;.\]
The maximum decreases with rising temperature, but its position 
remains at 1.9 eV.
Fig. 10 displays the contributions of the Fresnel- and
transmission factors to the SHG intensity. We find a strong
increase at energies larger than 2 eV,
due to the interband transition threshold in this energy
range. Decreasing the plasma frequency increases the SHG intensity and shifts 
the maximum to lower frequencies (e.g. for $\omega_{pl}=2.2$ eV we find a 
maximum at 1.1 eV, the intensity amplitude there is about $1000$ times larger).
The plasma frequency influences both Fresnel factors $f(\omega)$ 
and $F(2\omega)$. However since it is necessary that both 
$F(2\omega)$ and $f(\omega)$ are nonvanishing 
to make SHG possible, in the Fresnel factor contributions of the 
SHG yield mainly the effect of $f(\omega)$ is visible, 
because $F(2\omega)\neq 0$ for photon energies greater than energies 
where $f(\omega)$ is nonvanishing.

Analyzing the contributions of the linear and the nonlinear response
to the SHG intensity, we find that the 
overall shape of the intensity is dominated by the \cz-
contribution, if compared with the corresponding \cz-plots in
Fig. 9. The apparent shift of the \cz-plot to higher
photon energy results from the strong increase of the Fresnel factor 
contribution to the SHG in this energy range, as can be seen from
Fig. 10.

\begin{center}
\begin{table}
\begin{minipage}{12cm}
\caption{Fletcher - Wohlfahrt parameters fitted to {\em ab initio }
  calculations for the Cu monolayer.($E_0$: On-site energy, $A_i$:
  Overlap integrals of the $d$-orbitals)}
\end{minipage}
\begin{minipage}{8cm}
\begin{tabular}{|c|l|}
$\;\;\;\;\;\;\;$parameter$\;\;\;\;\;\;\;$
& $\!\!\!${\rm value } [eV] $\;\;\; \;\;\;$\\ \hline 
$E_0$ & $-2.904$ \\ \hline
$A_1$ & $0.106$ \\ \hline
$A_2$ & $0.088$ \\ \hline
$A_3$ & $0.147$ \\ \hline
$A_4$ & $0.220$ \\ \hline
$A_5$ & $0.170$ \\ \hline
$A_6$ & $0.200$ \\ \hline
\end{tabular}
\end{minipage}

\end{table}
\end{center}
\begin{figure}
\noindent
\caption[]{Change of the SHG intensity $\Delta I^{(2)}(\omega) = 
   \frac{I^{(2)}(\omega,T_{el})-
  I^{(2)}(\omega,300 K)}{I^{(2)}(\omega,300 K)} $ due to a 
rise of electron temperature as a function of the incident 
photon energy $\hbar \omega$ in percent for $T_{el} = 2000 K$ and
$T_{el} = 6000 K $.
The inset shows the energy range between 0.4 and 1.5 eV at an enlarged
abscissa scale. 
}
\end{figure}
\begin{figure}
\noindent
\caption[]{Change of the second order susceptibility 
 $ \Delta \chi^{(2)}(\omega  ,T) =\frac{\chi^{(2)}(\omega  ,T_{el}) -
\chi^{(2)}(\omega  ,300 K)}{ \chi^{(2)}(\omega  ,300 K)} $
due to a rise of electron temperature as a function of the incident 
photon energy $\hbar \omega $ in percent. 
$ \chi^{(2)}_{zzz}$ is very similar to the SHG yield for energies 
below 3.2 eV, indicating that the temperature effects for these 
fundamental energies on the SHG intensity
are caused by $ \chi^{(2)}_{zzz}$. The inset displays this similarity for 
incident energies between 0.4 and 1.5 eV.
}
\end{figure}
\begin{figure}
\noindent
\caption[]{Results for the relative change of the SHG yield $ \Delta
  \frac{I^{(2)}}{|\chi^{(2)}|^2}= 
 \frac{I^{(2)}(\omega ,T_{el})/{|\chi^{(2)}(\omega ,T_{el})|^2} -
I^{(2)}(\omega ,300 K)/ {|\chi^{(2)}(\omega ,300 K)|^2}} 
{ I^{(2)}(\omega ,300 K)/{|\chi^{(2)}(\omega ,300 K)|^2}} $
 for  $T_{el} = 2000 K$ and $T_{el} = 6000 K $.
In contrast to the results shown in Fig. 1, the frequency and temperature
dependence of \cwt is neglected. The inset shows at an enlarged
abscissa scale the energy range between 0.4 and 1.5 eV.
These results demonstrate the importance of the temperature dependence
of \cwt .
}
\end{figure}
\begin{figure}
\noindent
\caption[]{Contributions of the different electronic transition
  combinations to the SHG intensity as function of the incident photon energy 
$\hbar \omega$ (arbitrary units). 
Note, the contributions due to transitions with the incident state a $d$-band
(curve (b)) nearly coincides with the total SHG intensity.  
}
\end{figure}

\begin{figure}
\noindent
\caption{Calculated bandstructure of a Cu monolayer using the Combined
Interpolation Scheme. The $d$ band edge is 1.45 eV below
$E_F$. $\Delta$ is the interband transition onset energy. The
symmetry of the bands is indicated by the $s$ and $d$ labels.
}
\end{figure}

\begin{figure}
\noindent
\caption[]{Real and imaginary parts of the dielectric 
function as a function of the incident photon energy $\hbar \omega$ 
for three nonequilibrium electronic temperatures (300 K,
2000 K, 6000 K), calculated from Eq. (4). The inset shows the DOS of the Cu monolayer bandstructure.}
\end{figure}

\begin{figure}
\noindent
\caption[]{Illustration of important electronic transitions
  generating SH from noble metals. Note, both initial and intermediate
  state belong to the $d$-band. Actually, by neglecting such
  transitions in our calculation, we find a drastically reduced \iwt
  . Since in the case of the linear optical response, at best only the
  initial state can be a $d$ state, nonequilibrium temperature effects
  are drastically reduced and less interesting with regards to probing
  the electronic structure.}
\end{figure}
\begin{figure}
\noindent
\caption[]{Calculated total SHG intensity of a Cu monolayer 
for three different electronic
temperatures (300 K, 2000 K, 6000 K) as a function of the incident
photon energy $\hbar \omega$ (arbitrary units).
This quantity has been used for the calculation of the intensity changes. 
}
\end{figure}
\begin{figure}
\noindent
\caption[]{Calculated total susceptibility $ \chi^{(2)}_{zzz} $
 for three different 
temperatures (300 K, 2000 K, 6000 K) as a function of the incident 
photon energy $\hbar \omega$ (arbitrary units). 
These results have been used for the calculation of the susceptibility
changes given in Fig. 2 .
}
\end{figure}
\begin{figure}
\noindent
\caption[]{Temperature dependence
of the SHG intensity resulting from the temperature dependence of the
Fresnel factors and putting \cz equal to unity in Eq. (3). These results
have been used to calculate the differences in Fig. 3.

}
\end{figure}


\begin{thebibliography}{99}
\bibitem{hohlf95} J. Hohlfeld, D. Grosenick, U. Conrad, and
  E. Matthias, Appl. Phys. A {\bf 60}, 137 (1995)
\bibitem{stampfli90} P. Stampfli and K. H. Bennemann,
Phys. Rev. B {\bf 42}, 7163 (1990)
\bibitem{stampfli92} P. Stampfli and K. H. Bennemann,
Phys. Rev. B {\bf 46}, 10686 (1992)
\bibitem{hohlf95b} J. Hohlfeld, U. Conrad, and E. Matthias, to appear
\bibitem{hub_bohm94} W. H\"ubner, K. H. Bennemann, and K. B\"ohmer, 
Phys. Rev. B {\bf 50}, 17597 (1994)
\bibitem{klausdiss} K. B\"ohmer, Ph.D. thesis, Freie Universit\"at Berlin, 1994
\bibitem{sipe} J. E. Sipe, D. J. Moss, and H. M. van Driel, 
Phys. Rev. B {\bf 35}, 1129 (1987) 

\bibitem{ehren-hodges}H. Ehrenreich and L. Hodges, 
Methods Comput. Phys. {\bf 8}, 149 (1968)
\bibitem{fletcher-wohl} G. C. Fletcher and E. P. Wohlfahrt,
  Philos. Mag. {\bf 42}, 106 (1951) 
\bibitem{fletcher} G. C. Fletcher, Proc. Phys. Soc. (London)
{\bf A65}, 192 (1952)
\bibitem{freeman79} H. Krakauer, M. Posternak, and A. J. Freeman, 
Phys. Rev. B {\bf 19}, 1706 (1979).
\bibitem{weinb94} V. Drchal, J. Kudrnovsky, and P. Weinberger, 
Phys. Rev. B {\bf 50}, 7903 (1994).

\bibitem{haug84} H. Haug and S. Schmitt-Rink, 
Prog. Quant. Electr. {\bf 9} 3 (1984)
\bibitem{pusto93} U. Pustogowa, W. H\"ubner, and K. H. Bennemann, 
Phys. Rev. B {\bf 48}, 8607 (1993)
\bibitem{johnchrist72} P. B. Johnson and R. W. Christy,
Phys. Rev. B {\bf 6}, 4370 (1972).
\bibitem{hubner90} W. H\"ubner, 
Phys. Rev. B {\bf 42}, 11553 (1990)
\bibitem{footnote1}$\tau_{pl}$ cannot be calculated within RPA,
  contrary to $\omega_{pl}$.
\bibitem{hohlf95c} J. Hohlfeld and E. Matthias, private communication
\bibitem{hue_luce96} W. H\"ubner, T. A. Luce, K. H. Bennemann, to appear

\end{thebibliography}
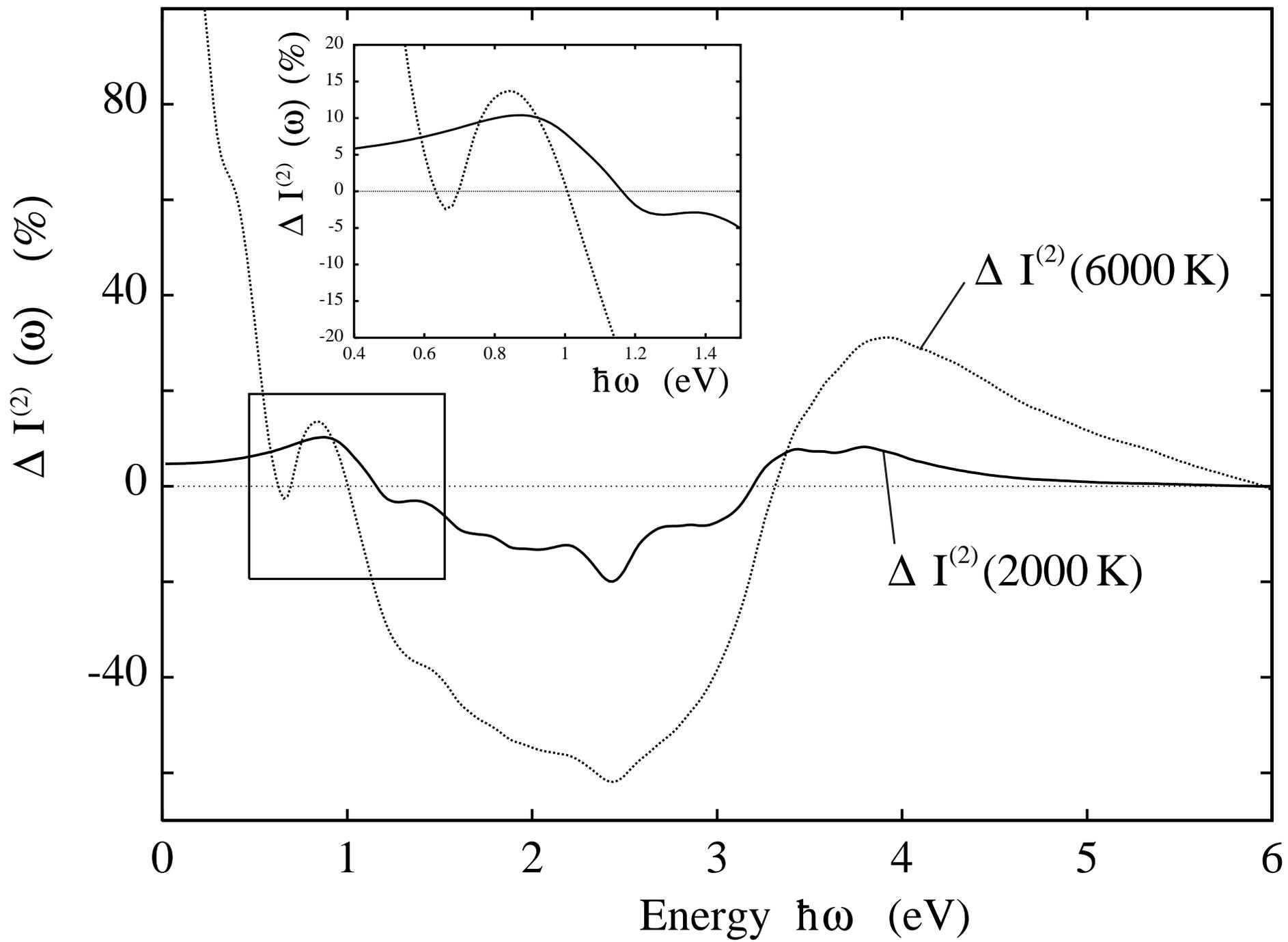
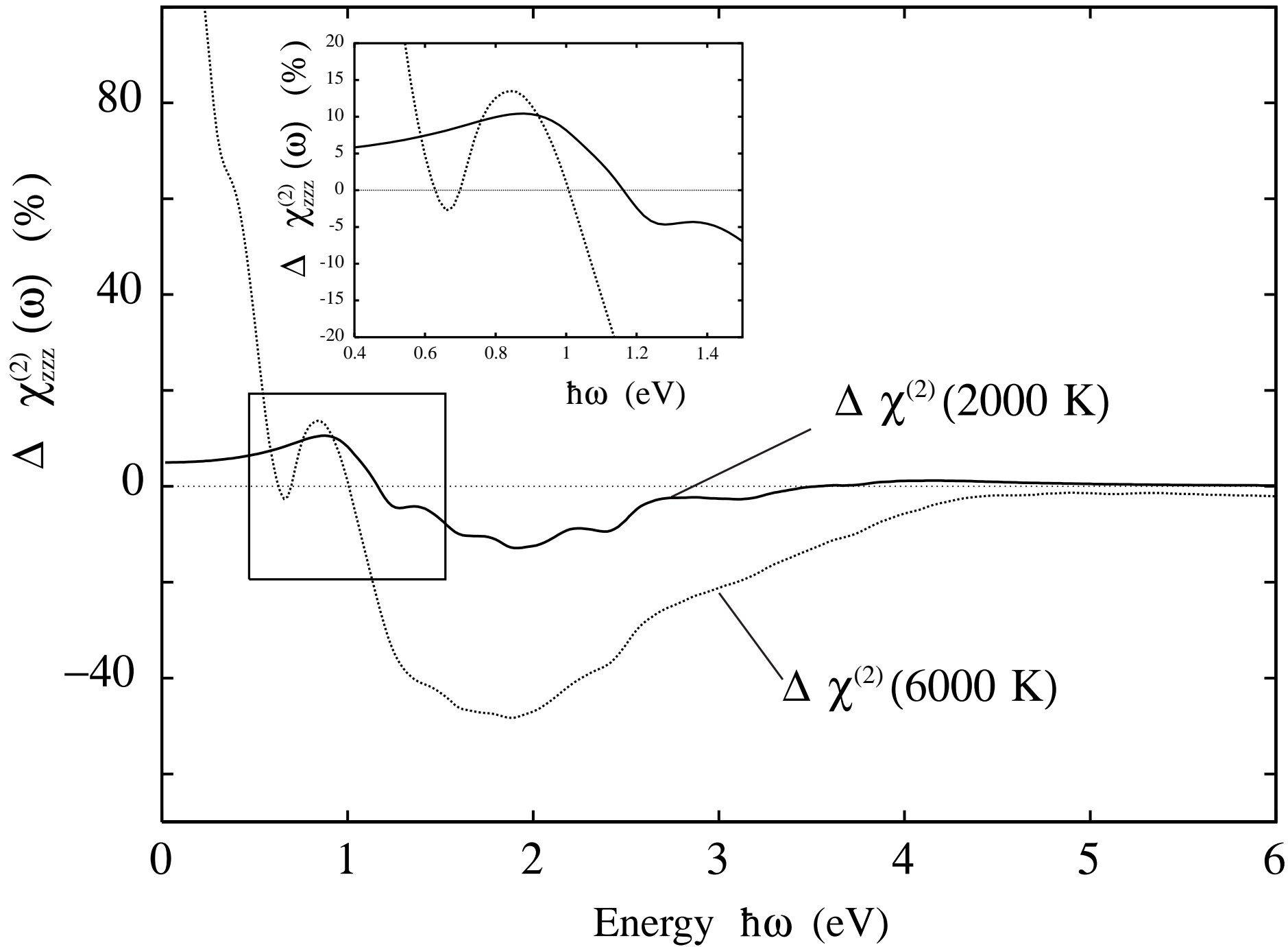
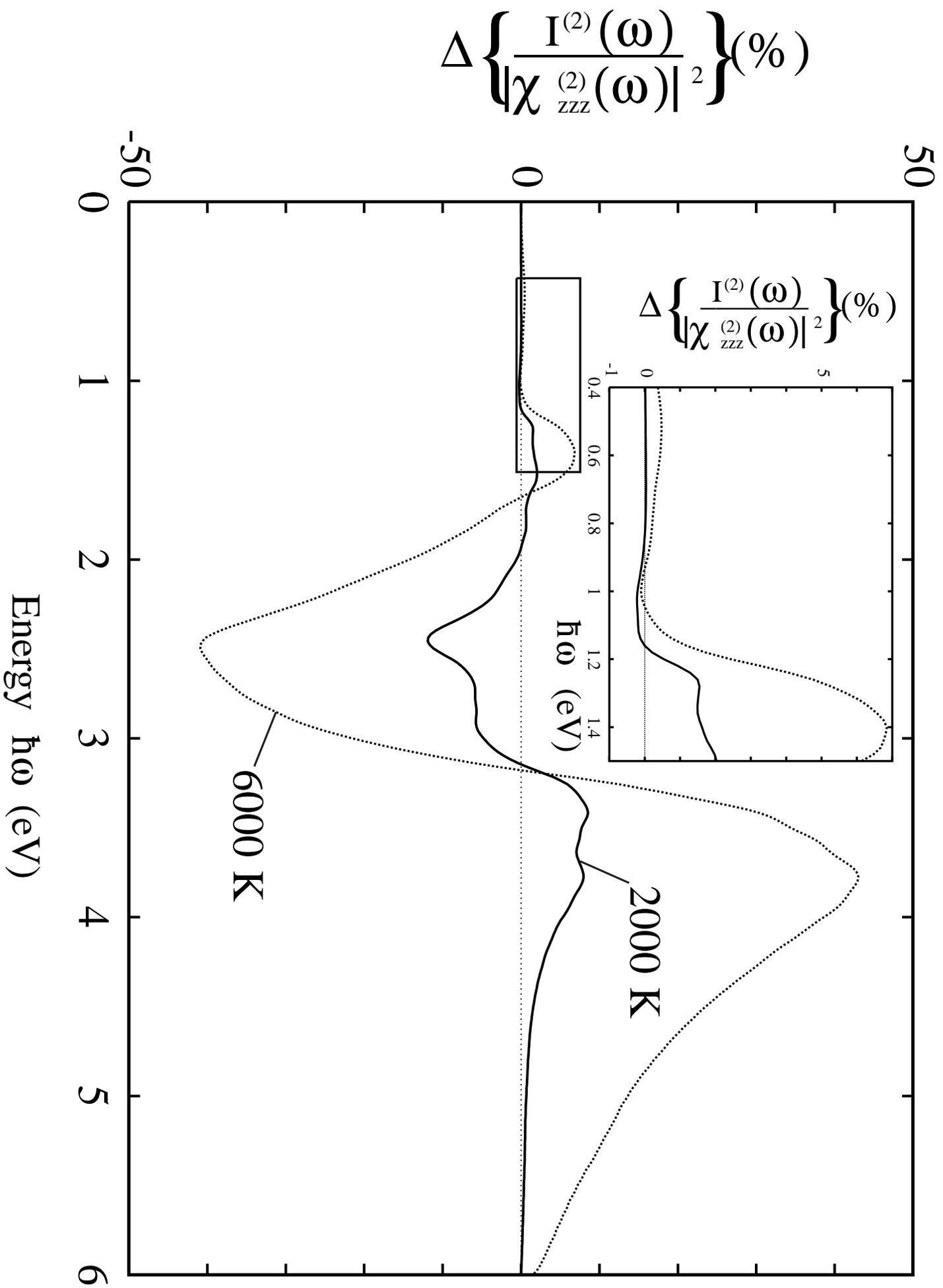
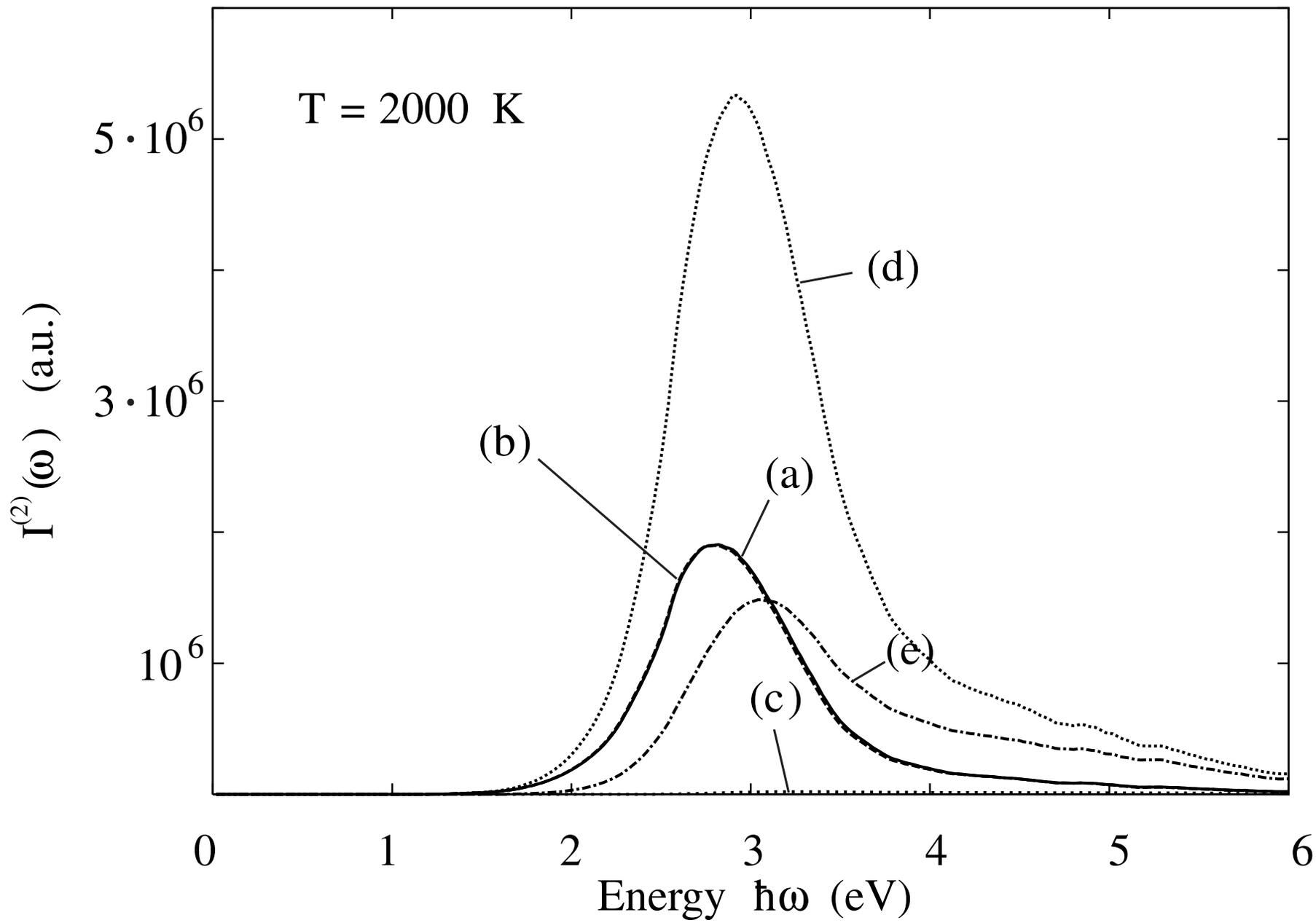
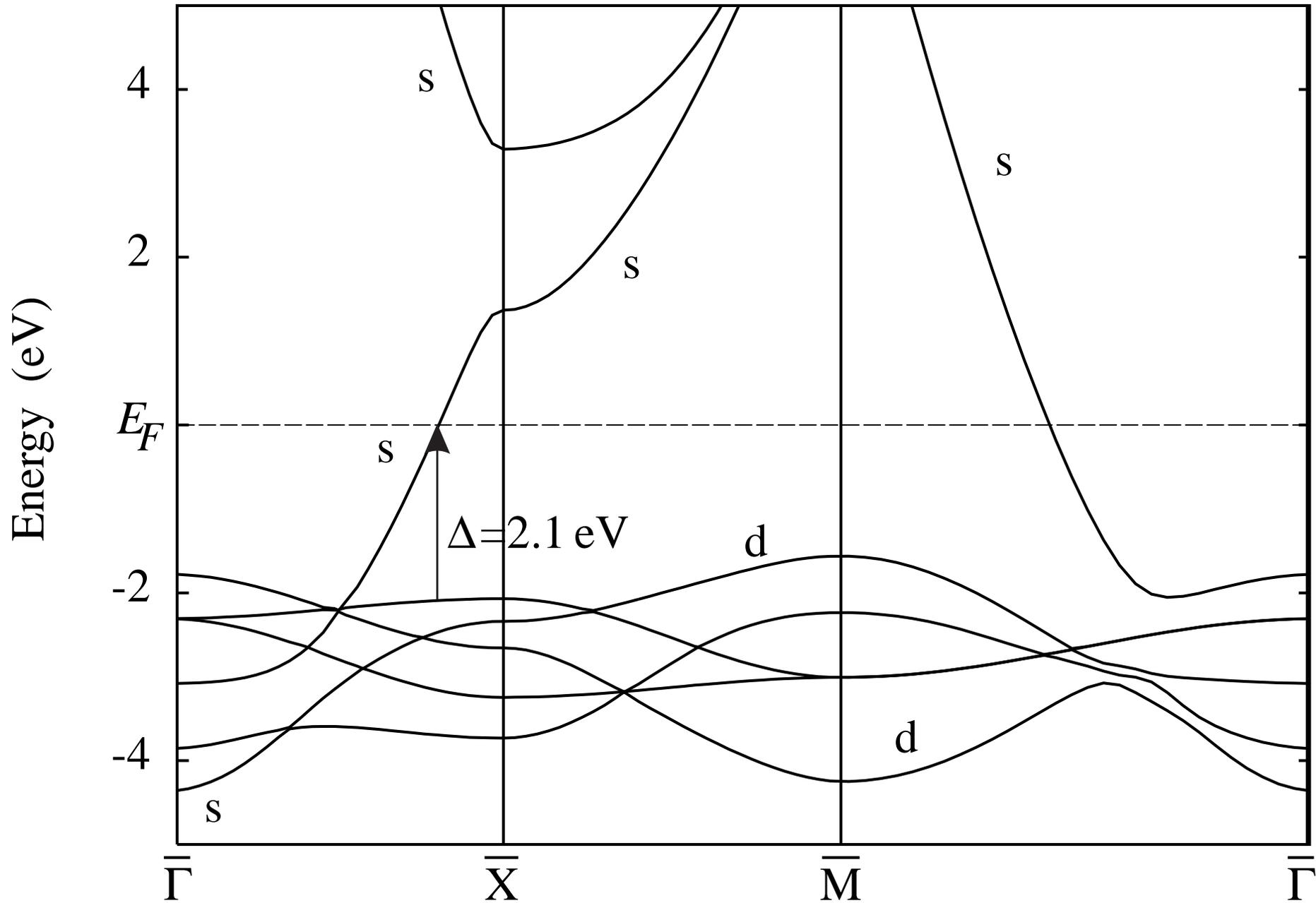
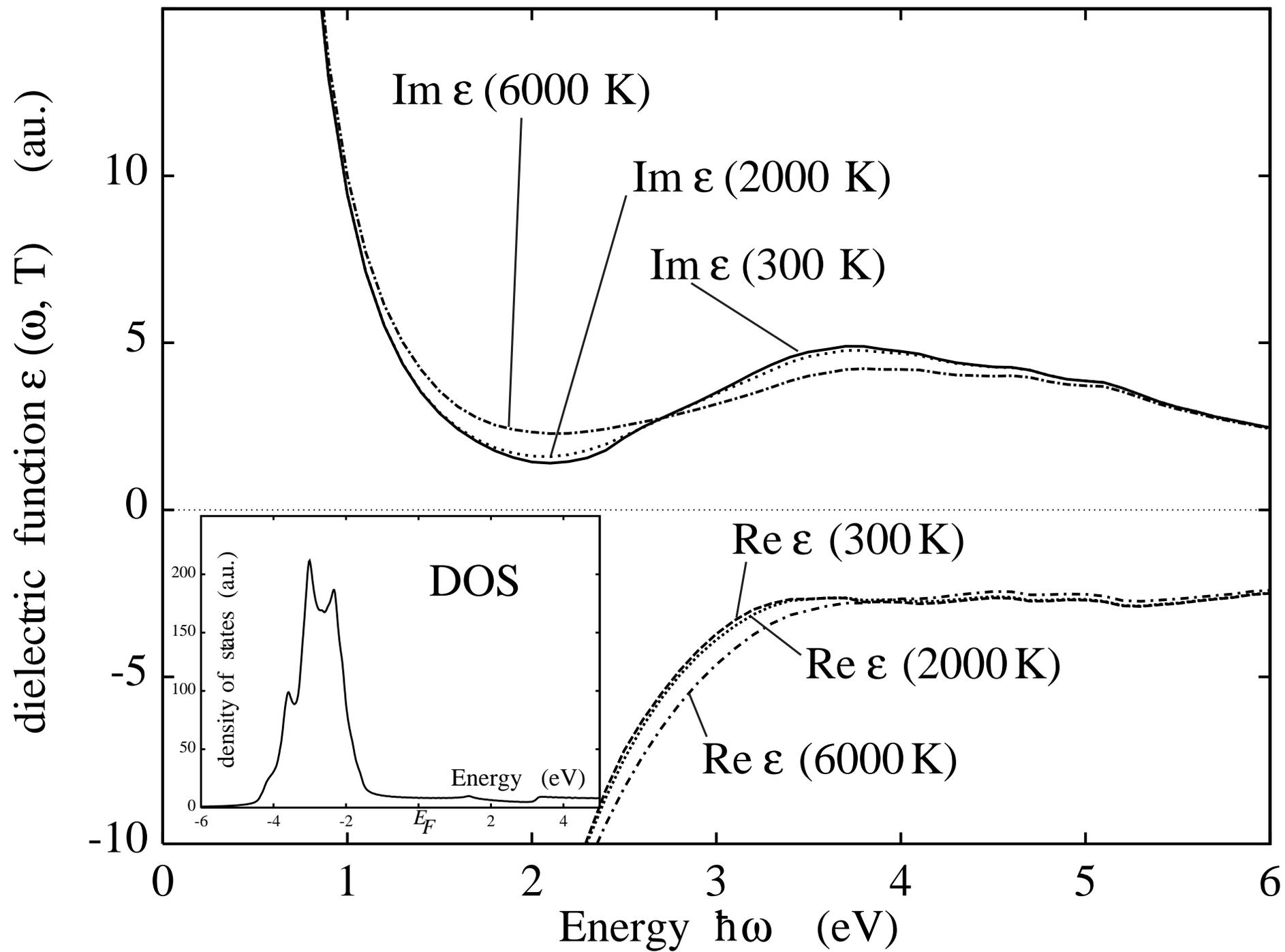
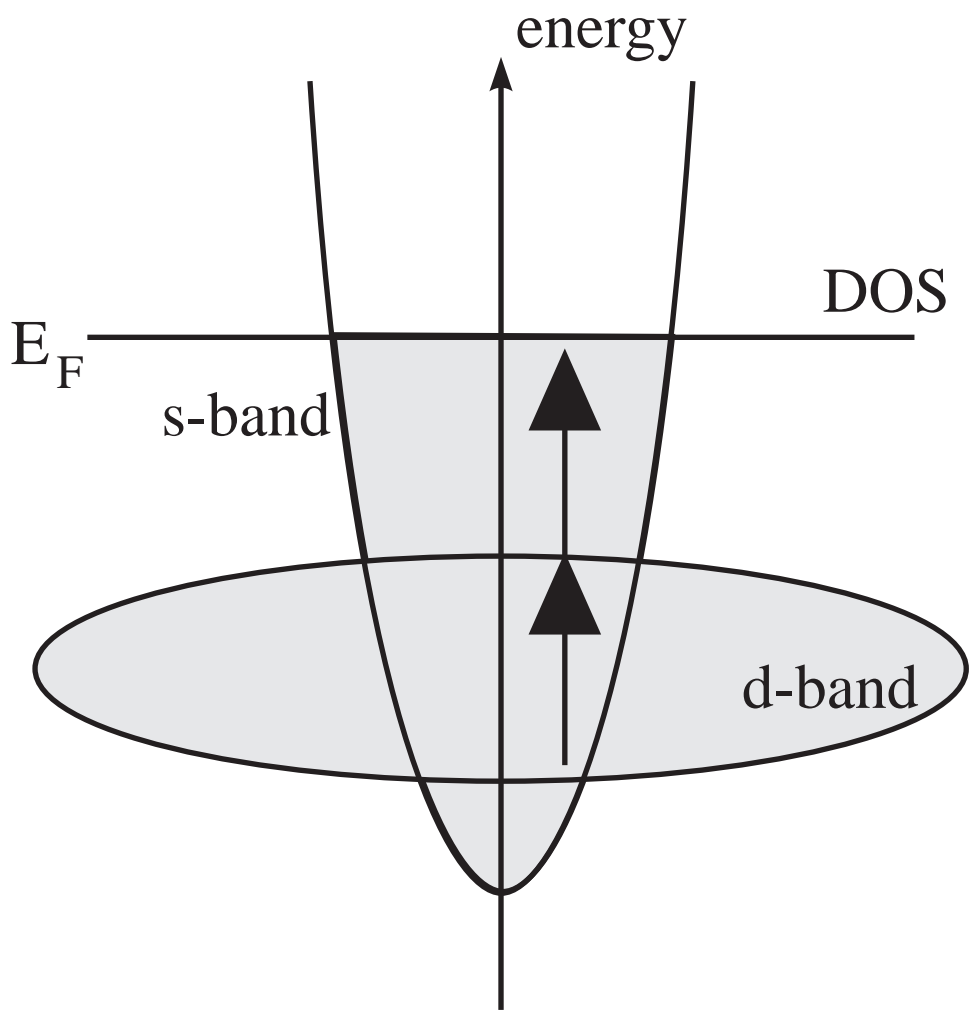
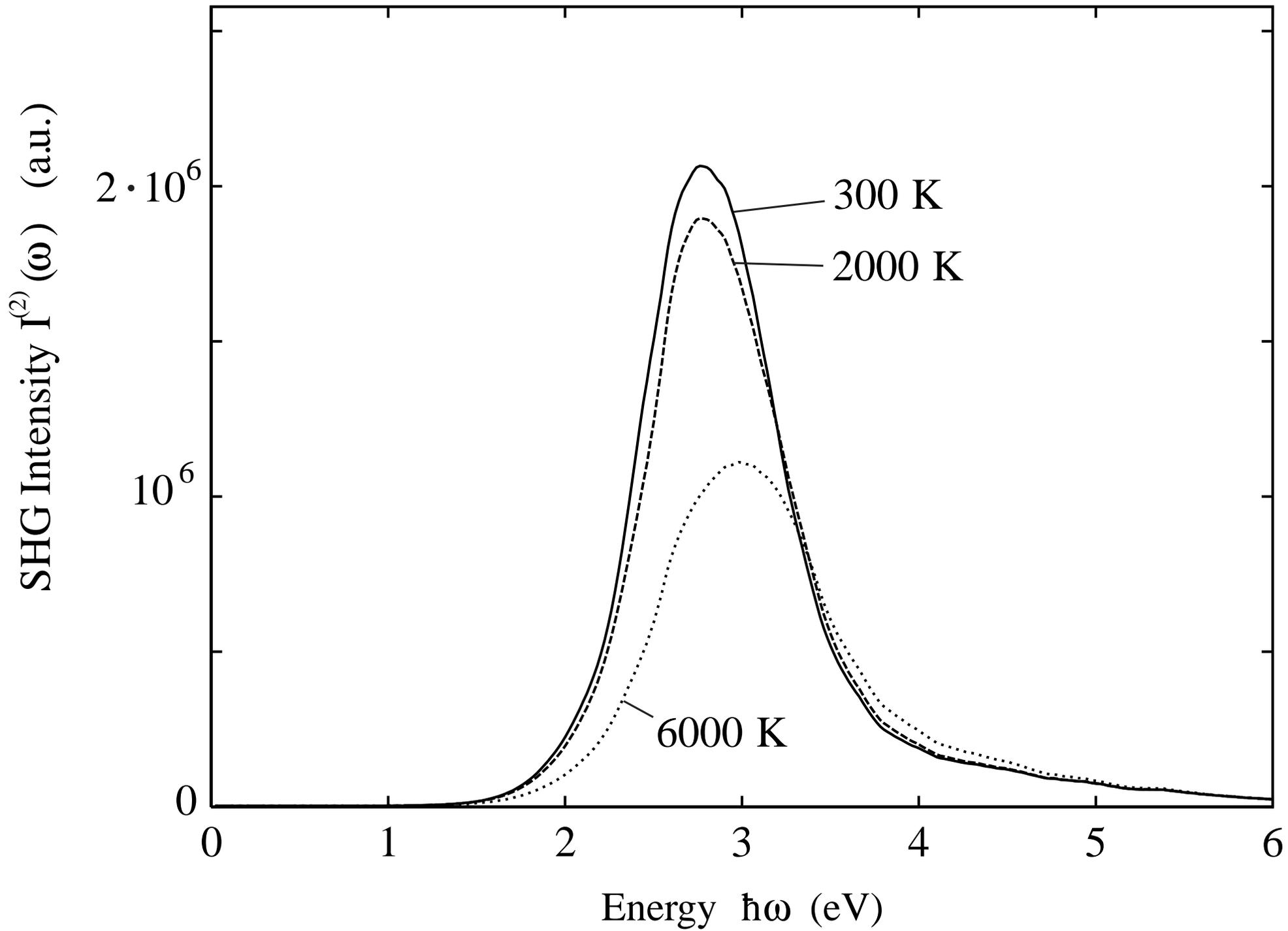
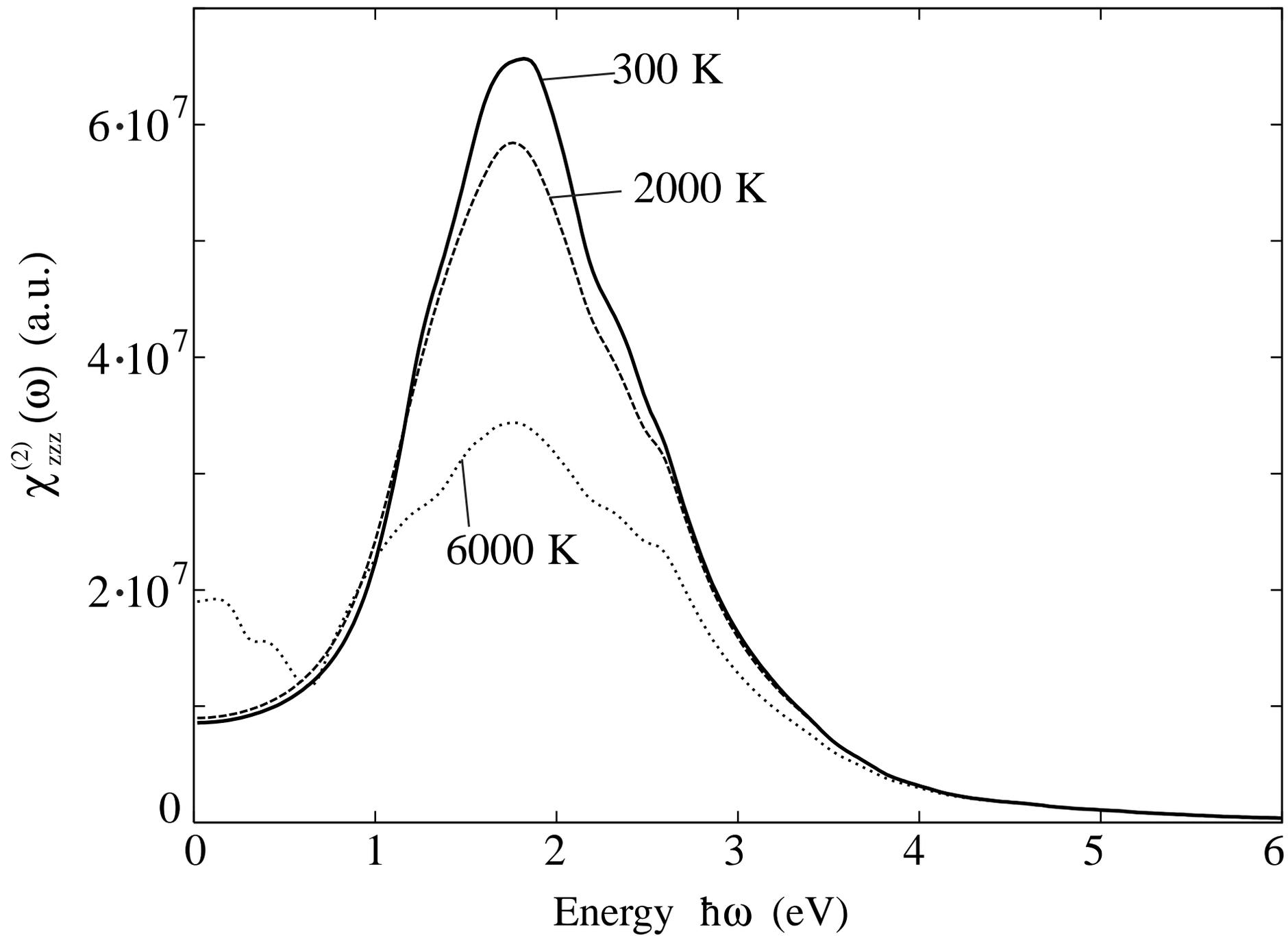
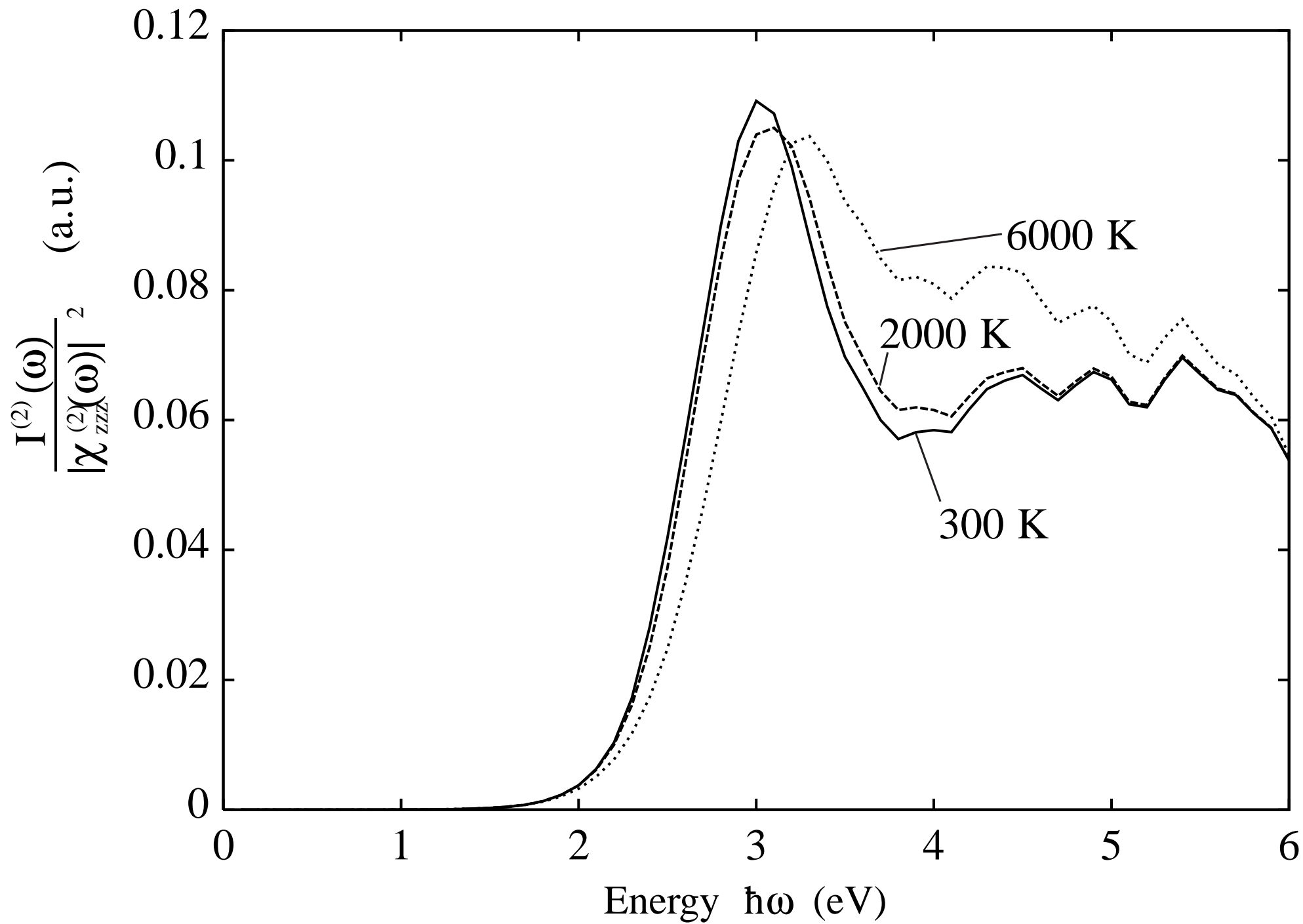
\end{document}